\documentclass[sigconf]{acmart}

\setcopyright{rightsretained}

\usepackage[font={small,sf}]{caption}
\usepackage{amsfonts}
\usepackage{setspace}
\usepackage{enumitem}
\usepackage{amsmath}
\usepackage{algorithmic}
\usepackage{algorithm}

\newcounter{myprot}
\newcounter{myalg}
\newcounter{mythm}
\newenvironment{mythm}
{\refstepcounter{mythm}  \noindent \textbf{THEOREM \arabic{mythm}:}\em}
{\vspace{.25em}}

\newcounter{mylem}
\newenvironment{mylem}
{\refstepcounter{mylem}  \noindent \textbf{LEMMA \arabic{mylem}:}\em}
{\vspace{.25em}}

\newcounter{mycor}
\newcounter{myobs}
\newcounter{mydef}
\newcounter{myconj}
\newenvironment{myproof}
{\noindent \textbf{PROOF:}}
{\vspace{-3ex}\begin{flushright} $\Box$ \end{flushright}\vspace{2ex}}

\usepackage{mathtools,url}
\usepackage{latexsym}

\makeatletter
\g@addto@macro{\UrlBreaks}{\UrlOrds}
\makeatother

\usepackage[framemethod=tikz]{mdframed}
  
\surroundwithmdframed[
  skipabove=1em,
  skipbelow=0ex,
  outerlinewidth=0.4pt,
  innerlinewidth=0.4pt,
  middlelinewidth=1pt,
  middlelinecolor=white,
  bottomline=false,topline=false,rightline=false]{mythm}

\surroundwithmdframed[
  skipabove=1em,
  skipbelow=0ex,
  outerlinewidth=0.4pt,
  innerlinewidth=0.4pt,
  middlelinewidth=1pt,
  middlelinecolor=white,
  bottomline=false,topline=false,rightline=false]{mylem}

\newif\ifinappendix% Default is \inappendixfalse
\let\oldappendix\appendix% Store \appendix
\renewcommand{\appendix}{% Update \appendix
  \oldappendix% Default \appendix
  \inappendixtrue% Set switch to true
}

\usepackage{booktabs}
\usepackage{hyperref}
\usepackage{xfrac}
\usepackage{wrapfig,multirow,relsize}
\usepackage{microtype}

\newcommand{\OmitText}[1]{ {} }

\newcommand{\bobtail}{Bobtail\xspace}
\usepackage{xspace}

\newcommand{\para }[1]{\smallskip \noindent {\bf #1}}
\newcommand{\1}{{\em (i)}}
\newcommand{\2}{{\em (ii)}}
\newcommand{\3}{{\em (iii)}}
\newcommand{\4}{{\em (iv)}}

\usepackage[all]{hypcap}

\usepackage[capitalise,nameinlink]{cleveref}
\crefname{section}{Sect.}{Sect.}
\Crefname{section}{Section}{Sections}

\usepackage{url}
\makeatletter
\g@addto@macro{\UrlBreaks}{\UrlOrds}
\makeatother

\renewcommand\footnotetextcopyrightpermission[1]{} % removes footnote with conference information in first column
\title{\bobtail: Improved Blockchain Security\\ with Low-Variance Mining}

\author{George Bissias  \hspace{.6cm}Brian N.~Levine  \\
College of Information and Computer Sciences,  UMass Amherst}
\begin{document}

\begin{abstract}
\emph{Blockchain systems are designed to produce blocks at a constant average rate. The most popular systems currently employ a \emph{Proof of Work (PoW)} algorithm as a means of creating these blocks.  An unfortunate limitation of all deployed PoW blockchain systems is that the time between blocks has high variance. For example, Bitcoin produces, on average, one block every 10 minutes. However, 5\% of the time, Bitcoin's inter-block time is at least 40 minutes. }

\emph{In this paper, we show that high variance is at the root of several fundamental attacks on PoW blockchains. We propose an alternative process for PoW-based block discovery that results in an inter-block time with significantly lower variance.  Our algorithm, called \emph{Bobtail},  generalizes the current algorithm by comparing the mean of the \mbox{$k$-lowest} order statistics to a target. We show that the variance of inter-block times decreases as $k$ increases.  Bobtail significantly thwarts  doublespend   and selfish mining attacks, and makes detection of eclipse attacks trivial and quick. For example, for Bitcoin and Ethereum, a doublespending attacker with 40\% of the mining power will succeed with 53\% probability when the merchant sets up an embargo of 1~block; however, when $k\geq40$, the probability of success for the same attacker falls to less than 1\%.  Similarly, for Bitcoin and Ethereum currently, a selfish miner with 49\% of the mining power will claim about 95\% of blocks; however, when $k\geq20$, the same miner will find that selfish mining is less successful than honest mining. We also investigate attacks newly made possible by Bobtail and show how they can be defeated. The primary costs of our approach are larger blocks and increased network traffic.}
\end{abstract}

\maketitle

\urlstyle{sf}
\pagestyle{plain}

\section{Introduction}
\label{sec:introduction}

Blockchain systems are designed to produce blocks at a constant average rate. The most popular systems currently employ a {\em Proof of Work (PoW)} algorithm as a means of creating these blocks~\cite{Nakamoto:2009}, including Bitcoin Cash,  Bitcoin Core, Ethereum~\cite{ethereum}, and  Litecoin~\cite{litecoin}.
For example, Bitcoin and Bitcoin Cash produce one block every 10 minutes on average and  self-adjust the difficulty of producing a block  if too many or too few have been produced. 
Unfortunately, a limitation of all deployed PoW blockchain systems is that the time between blocks has high variance and the distribution of inter-block times has a very long tail. For example, 5\% of the time, Bitcoin's inter-block time is at least 40 minutes.  As we show, high variance enables  doublespend~\cite{Nakamoto:2009,Ozisik:2017} and selfish mining~\cite{Eyal:2014} attacks, in addition to impeding the consistent flow of validated transactions through the system.

The high variance of inter-block times is a direct consequence of  PoW algorithms. Generally, miners in  all deployed systems  craft blocks by repeatedly changing a nonce in the block header until the hash of that header is less than a target value $t$.  In other words, the hash of each header is a sample taken randomly from a discrete uniform distribution that ranges between $[0,S]$, where $S = 2^{b}-1$ and typically $b=256$. A block is discovered when the {\em first order statistic} (i.e., the minimum value) of all sampled values is less than target $t$, $0<t<S$.  

In this paper, we show that high variance is at the root of several fundamental attacks on PoW blockchains, including doublespend and selfish mining attacks. We propose an alternative process for PoW-based block discovery that results in an inter-block time with significantly lower variance.  Our algorithm generalizes the current algorithm by comparing the mean of the $k$-lowest order statistics to a target. We show that the variance of inter-block times decreases as $k$ increases.  For example, if our approach were applied to Bitcoin, nearly every block would be found within 5 to 18 minutes; and the average inter-block time would remain at 10 minutes. As a result, doublespend and selfish mining attack efficacies are drastically reduced. The cost of our approach is larger blocks and increased network traffic. New attacks are possible, but we show how they can be thwarted. 
We call our approach \emph{\bobtail}
mining. 
 
\para{Problem Statement.}

Imagine that the mining process is carried out for exactly $h$ hashes during time interval $I$, generating hash values $H_1, \ldots, H_h$ from $[0, S]$ uniformly at random. Now define $V_i$ to be the value of the $i$th order statistic at the end of $I$, i.e. $V_i = H_{(i)}$ in standard notation. Let $W_k$ be a random variable representing the average value over the lowest $k$ order statistics after $h$ hashes. 
\begin{equation}
W_k = \frac{1}{k} \sum_{i=1}^k V_i.\label{eq:wk}
\end{equation}

$W_k$ constitutes the collective mining proof ({\em proof}, for short) for the entire network. Our \bobtail mining criterion says that a new block is discovered   when a realized value of $W_k$ meets the target $t$:
\begin{equation}
w_k \leq t. \label{eq:btmc}
\end{equation}
Notably, this approach is a generalization of current systems, which are the special case where $k=1$. 

Our primary goals are therefore to show, given values of $k>1$, that: \1
there is a significantly reduced inter-block time variance;  \2 the costs are relatively small, which include increases in block size and network traffic; \3 selfish mining and doublespend attacks are significantly more difficult to carry out as $k$ increases; and \4 that new attacks made possible by setting $k>1$ are easily mitigated.

\para{Contributions.} 
\begin{itemize}
\item We derive the statistical characteristics of our approach and validate each empirically. For example, we derive expressions for the expectation and variance of the \bobtail mining proof and the number of hashes performed for any value of $k$. Using these expressions, we quantify the reduction in variance of inter-block time expected for values of $k>1$. 
\item We show that variance in block discovery time is reduced by $O(1/i)$ when using $k = i$ order statistics relative to the variance when $k = 1$ (the status quo). 
\item We show that  \bobtail mining results in a lower rate of  {\em orphaned blocks}. 
\item We demonstrate that low-variance mining significantly mitigates the threats to security posed by selfish mining and doublespend attacks. 
 For Bitcoin and Ethereum currently (i.e., when $k=1$), an attacker with 40\% of the mining power will succeed with 30\% probability when the merchant sets up an embargo of 8 blocks; however, when $k\geq20$, the probability of success falls to less than 1\%.  Similarly, for Bitcoin and Ethereum currently, a selfish miner with 45\% of the mining power will claim about 95\% of blocks; however, when $k\geq20$, the same miner will find that selfish mining is less successful than honest mining. 
 
\item We show that new intra-block withholding attacks are possible for Bobtail. However, by carefully designing a rewards scheme for mining, these attacks are mitigated. 

\end{itemize}

\section{Related Work}

Our approach is related to previous results in proof-of-work, cryptographic puzzles, and improvements to blockchain architectures. 

\para{Proof of work.} A large number of papers have explored applications of proof-of-work. Dwork and Naor~\cite{Dwork:1992} first suggested proof-of-work (PoW) in 1992,  applying it as a method to thwart spam email. A number of subsequent works similarly applied PoW  to thwarting denial-of-service (DoS) attacks~\cite{Aura:2001,Groza:2014,Dean:2001,Wang:2003,Franklin:1997,Chen:2001}. Our approach can be adopted into many of these past works to improve computational variance. Jakobsson and Juels~\cite{Jakobsson:1999} and Jules and Brainerd~\cite{Juels:1999} examine the security properties of PoW protocols, and base their theorems on the average work required; our approach would provide stronger guarantees under their theorems since the variance is lower.  Laurie and Clayton~\cite{Laurie:2004} examine the  practical limitations in deploying PoW solutions in DoS scenarios.

Douceur~\cite{Douceur:2002} noted in 2002 that proofs of work can mitigate Sybil attacks. Also in 2002, Back~\cite{Back:2002}  applied PoW to cryptocurrencies. Back noted the high variance of computational PoW and regarded it as an open problem. Nakamoto's Bitcoin~\cite{Nakamoto:2009} built on these ideas. 

In 2003, Abadi et al.~\cite{Abadi:2005} suggested memory-bound functions as a better foundation for avoiding the variance in CPU resources among users. Indeed,  the ETHASH~\cite{ethash} PoW algorithm in Ethereum~\cite{ethereum} adopted a PoW function that requires more memory than is economically profitable for custom ASICSs. In contrast, our goal is to reduce the variance of the entire network's time to solve a PoW problem, and it is not to increase egalitarianism or increase participation by eschewing specialized hardware. In any case, our approach is applicable to ETHASH.

Coelho~\cite{Coelho:2008} is the work is closest  to ours in terms of goals. That work proposes a PoW puzzle based on Merkle trees that requires an exact number of steps and therefore has no computational variance. Without variance, the same miner would always win, and therefore the method is unsuitable for blockchains.

Bitcoin-NG~\cite{Eyal:2016} also offers low-variance transaction announcements via PoW-based leader election. However, because inter-leader time variance would still be from an exponential distribution, unlike Bobtail, Bitcoin-NG does not reduce doublespend, selfish mining, or eclipse attack vulnerability. Furthermore, unlike our approach, Bitcoin-NG sets up the elected leader as a single point of failure.

\para{Blockchains without PoW.} Several newer blockchains are not based on computational proof of work, and our solution does not apply to them. These include proof-of-storage~\cite{Miller:2014}, proof-of-stake~\cite{Bentov:2016,Bentov:2014}, and blockless~\cite{Boyen:2016} schemes. However, almost all wealth stored in cryptocurrencies are in computational PoW blockchains that our approach does apply to, including Bitcoin, Bitcoin Cash, Litecoin, Ethereum, and Ethereum Classic.  

\section{Protocol}
\label{sec:protocol}

In this section, we 
provide  a high-level overview of the Bobtail protocol and its fundamental features. 

\begin{figure}[t]
   \centering
   \includegraphics[width=\columnwidth]{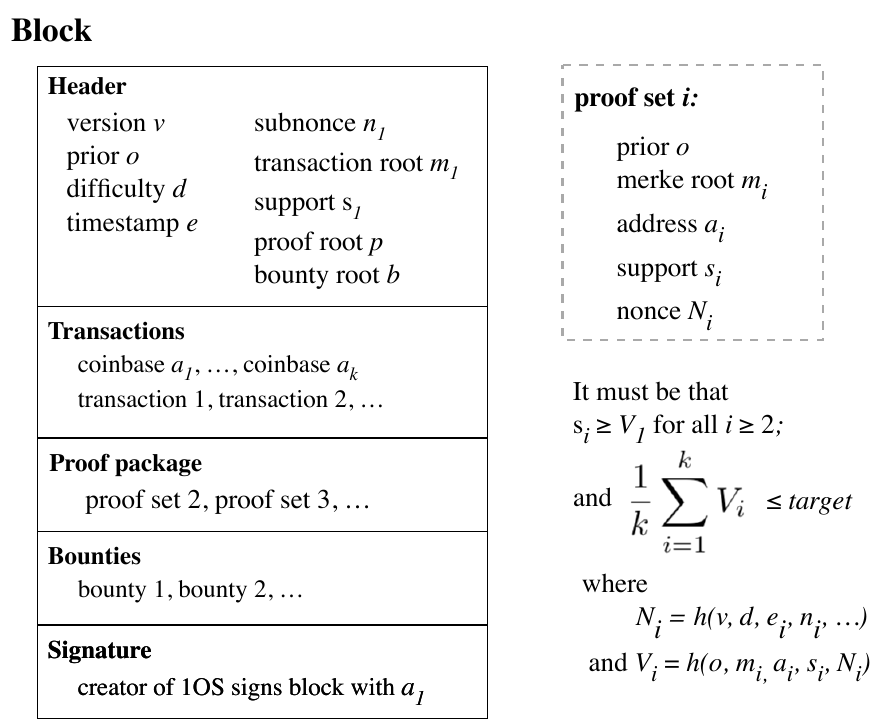} 
   \caption{\bobtail blocks are a superset of existing PoW schemes. They contain a link to a prior block, a timestamp, the current difficulty, and the Merkle root of a set of transactions; they also add {\em proof sets} contributed by other miners as well as \emph{bounties} that prove claims about individual proof sets.}
   \label{fig:overview}
\end{figure}

\para{Blocks.} Bobtail blocks, illustrated in Figure~\ref{fig:overview}, consist of a {\em block header} $\mathcal{H}$, a set $\mathcal{T}_1$ of valid, previously unconfirmed {\em transactions}, a {\em proof package} $\mathcal{K}$, bounties $\mathcal{B}$, and a signature $\mathcal{S}$. Block header $\mathcal{H}$ contains the same fields as a conventional block header such as the one used in Bitcoin or Ethereum, and three additional fields, which are described below. Proof package $\mathcal{K}$ is a collection of $k$ {\em proof sets}, where each proof set $\mathcal{P}_i$ contains a payout address for the miner, values necessary for creating valid proof of work, and other values used for thwarting attacks. Proof sets are hashed to create PoW, i.e. $h(\mathcal{P}_i) = V_i$. The sets are ranked so that $V_1$ is defined as the smallest value or \emph{first order statistic} (1OS, for short). Each bounty $\mathcal{B}(T, \mathcal{T}) \in \mathcal{B}$ is a Merkle proof that some transaction $T$ is in $\mathcal{T}$ for some transaction set $\mathcal{T} \in \{\mathcal{T}_1, \ldots, \mathcal{T}_k\}$; it is used to prevent doublespend attacks as we describe in detail below. Finally, $\mathcal{S}$ is the signature of $\mathcal{H}$ and it must be generated with the private key that matches the payout address in $\mathcal{P}_i$. 

\para{Mining.} Bobtail mining is a generalization of the procedure implemented in conventional PoW blockchains. Each miner seeks to receive coinbase reward by becoming one of the $k$ proof sets included in $\mathcal{K} = [\mathcal{P}_1, \ldots, \mathcal{P}_k]$. Proof set $\mathcal{P}_i $ has fields $(o, m_i, a_i, s_i, N_i)$, where $o$ is the hash of the previous block header, $m_i$ is the root of a Merkle tree containing transactions $\mathcal{T}_i$, $a_i$ is the miner's payout address, $s_i$ is the supporting proof for $\mathcal{P}_i$, and $N_i = h(\mathcal{N}_i)$, the nonce. In contrast to typical PoW systems, the nonce is more than just a random number; it contains values that can be omitted from the blockchain for all proofs except the 1OS. Specifically, $\mathcal{N}_i = (v, d, e_i, n_i, \ldots)$, where $v$ is protocol version, $d$ is current difficulty, $e_i$ is a timestamp, $n_i$ is a nonce, and we allow zero or more optional arguments. 
Finally, header $\mathcal{H}$ has fields $(v, o, d, e, N_1, m_1, s_1, p, b)$, where $e$ is a timestamp, $p$ is the root of a Merkle tree containing every $V_i$ corresponding to $\mathcal{P}_i \in \mathcal{K}$, $b$ is the root of a Merkle tree containing $h(\mathcal{B}(\mathcal{T}, T))$ for each $\mathcal{B}(\mathcal{T}, T) \in \mathcal{B}$, and the remaining fields match the corresponding values from $\mathcal{P}_1$.

Like other PoW blockchains, miners select new nonces and generate proofs continuously. In Section~\ref{sec:overhead}, we show how a miner can precisely determine the probability that any given proof will eventually be included in the mining package. Once a proof is discovered having sufficient probability of inclusion, the corresponding values $\mathcal{H}$, $\mathcal{P}_i$, and $\mathcal{T}_i$ are propagated. Because of their large size and high degree of redundancy, transaction sets $\mathcal{T}_i$ can be propagated using Graphene~\cite{Ozisik:2017c} or the weak block protocol~\cite{Rizun:2016} to greatly reduce associated network overhead. 
 
The difficulty $d$ for each block is adjusted roughly once every two weeks using the same algorithm deployed in Bitcoin\footnote{\url{https://github.com/bitcoin/bitcoin/blob/78dae8caccd82cfbfd76557f1fb7d7557c7b5edb/src/pow.cpp\#L49}}. In short, the mean block time for the last 2016 blocks is used to estimate the actual difficulty at which the miners were operating; then the difficulty is adjusted up or down in order to ensure that the expected block time is 10 minutes if miners continuing mining at the same rate. At a given difficulty $d$, the target $t$ is derived from $d$ in the same manner that it is for Bitcoin\footnote{\url{https://github.com/bitcoin/bitcoin/blob/78dae8caccd82cfbfd76557f1fb7d7557c7b5edb/src/pow.cpp\#L80}}, which involves translating integer $d$ into a threshold 256-bit arithmetic value (i.e., one that supports arithmetic operations). We say that a block is \emph{assembled} when \1 the mean of the $k$ proofs, $V_1, \ldots V_k$, is less than or equal to $t$; \2 the package is signed by the miner who generated the 1OS using address $a_1$; \3 supports $s_2,\ldots s_k$ are greater than or equal to $V_1$. 

The coinbase transaction of the block creates new coin that is distributed to addresses $a_1, \ldots, a_k$ according to the scheme described in Section~\ref{sec:rewards}. 
If the coinbase of the new block does not contain the correctly awarded coinbase, then the block is ignored by other miners.
The assembled block, comprised of $\mathcal{H}$, $\mathcal{T}_1$, $\mathcal{K}$, $\mathcal{B}$, and $\mathcal{S}$ is propagated throughout the network. Receivers validate that $\mathcal{S}$ was generated by hashing $\mathcal{H}$ with the private key corresponding to payout address $a_1$. They also validate $\mathcal{T}_1$, $\mathcal{K}$,  and $\mathcal{B}$ against $\mathcal{H}$. If validation succeeds, then the block is added to the blockchain. 

\para{Bounties} Bobtail employs a \emph{bounty} system to disincentivize the use of incompatible transactions (those spending the same UTXO) in transactions sets $\mathcal{T}_i$. Bounty $\mathcal{B}(\mathcal{T}_i, T')$ contains a Merkle proof that transaction set $\mathcal{T}_i$ contains transaction $T'$. The Merkle proof is simply all nodes along a path from the Merkle root $m_i$ to the leaf containing $T'$. When proof set $\mathcal{P}_i$ is disseminated in the network, it must be accompanied by Merkle root $m_i$ and associated transactions $\mathcal{T}_i$. The recipient checks that the transactions in $\mathcal{T}_i$ are compatible with the transactions in her own set $\mathcal{T}_j$. If she discovers incompatible transaction $T' \in \mathcal{T}_i$, then she can create bounty $\mathcal{B}(\mathcal{T}_i, T')$. Later, if one of her proofs is the 1OS, she will include all bounties that implicate transactions in conflict with $\mathcal{T}_1$. 

\para{Fork-choice rule.} If multiple miners generate proofs with value low enough to mine a block with the same parent, then there can arise ambiguity over which extends the \emph{main chain}, i.e. the chain that honest miners will continue to extend. To avoid this ambiguity, we define the main chain to be the one comprising the most \emph{aggregate work}, from the genesis block up to the tip; all competing chains are \emph{orphaned}, i.e. ignored by honest miners. Aggregate work is calculated as the sum of inferred hashes, $S / w_k$, over each block, where $S$ is the size of the hash space and $w_k$ is the value of the mining statistic. In general, this fork-choice rule ensure that blocks with lower average proof values will be favored over those with higher values. Note that because proof sets reference a specific parent block, they can only be shared between child blocks having the same parent. And according to our fork-choice rule, ultimately the child block with the lowest value $w_k$ will extend the main chain. 

\para{Additional rules.} In order to reduce the number of orphaned blocks (discussed in Section~\ref{sec:orphan_prevention}) and thwart various attacks (see Section~\ref{sec:attacks}), miner $M$ will adhere to the following rules. \1 $M$ rejects proof package $\mathcal{K}$ if $V_1$ is higher than the lowest proof she has seen announced on the network. \2 When assembling a proof package as author of the 1OS, $M$ will include proofs from other miners in the same order she received them from the network. Specifically, $M$ begins by identifying all sets of $k$ proofs $S_1, \ldots$, each with mean value below $t$. Let $r$ be the maximum reward value, across all $S_i$, that would be returned to $M$ if the the given set was assembled into a block. She discards any sets that do not return reward value $r$, and then assembles the proof package from the remaining set with the earliest average proof receipt time.

\section{Threat Model}
\label{sec:attacks}

\para{Doublespend and selfish mining.} Doublespend~\cite{Nakamoto:2009,Ozisik:2017} and selfing mining~\cite{Eyal:2014} attacks are the two most fundamental attacks on blockchains~\cite{Sapirshtein:2015,Gervais:2016}. In both, attackers attempt to mine a fork of the blockchain that is longer than the honest miners' branch. Because the attacker is assumed to have a minority of the mining power, in expectation, the attacker cannot create a longer branch than that of the honest miners.  However, just like a person visiting a casino, the attacker is seeking a short-term win. He is attempting to get lucky and find a series of blocks quickly while the honest miners are relatively unlucky and discover blocks slowly, despite the larger amount of mining power. Intuitively the success of the attacks lies in leveraging the inherent variance of mining. 

\para{Proof withholding.} A proof withholding attack is unique to Bobtail and involves miner $A$ declining to publish some subset of her proofs immediately after they are generated. Instead, $A$ withholds the proofs in order to gain an informational advantage over the remaining miners, $M$. While $A$ sees all proofs, an honest miner in $M$ sees only proofs generated by members of $M$. $A$ hopes that this advantage will allow her to assemble some proof packages with more than her fair share of proofs and ultimately lead to an increase in her total reward.

\para{Zero-cost zero-confirmation (ZCZC) doublespend.} When carrying out a zero-confirmation doublespend attack, the attacker seeks to displace payment transaction $T$ with an incompatible transaction $T'$, which spends the same UTXO. The ZCZC doublespend is a variant of the conventional attack where the attacker \emph{reuses} proofs between multiple branches of the blockchain to this end. To do so, he releases payment transaction $T$ to a merchant and the Bitcoin network at large. Simultaneously he mines with transaction set $\mathcal{T}_j$ containing $T'$. If he is lucky enough to mine the 1OS, then $j=1$, $\mathcal{T}_j$ becomes the canonical transaction set for the block, $T'$ is confirmed, and the merchant payment is negated. But even if he does not mine the 1OS, he can release his proofs containing $\mathcal{T}_j$ at some point before the block is mined and still receive a reward for those that are included in the proof package. Note that this attack also applies to doublespends where the attacker attempts to produce a competing branch of arbitrary length $z$, however, proofs can only be reused during the creation of the first block on the competing branch.

\para{Denial-of-reward.} It is possible for an attacking peer in the network to cheaply weaponize the bounty process against miners. Suppose that peer $A$ creates incompatible transactions $T$ and $T'$ such that both spend the same UTXO. Suppose further that she disseminates $T$ to roughly half the miners and $T'$ to the remaining miners. Then, it is likely that the first group of miners will regard $T$ as a legitimate transaction and $T'$ as a doublespend, while the opposite would be true for the other miners. Thus, in expectation, roughly half the proofs will use a transaction set containing $T$ and half will use a transaction set containing $T'$. Furthermore, miners will construct bounties for whichever transaction they believe to be a doublespend. Eventually, a proof package will be assembled (containing transaction set $\mathcal{T}_1$) by one of the miners $M$, who holds the 1OS. Assume, without loss of generality, that $\mathcal{T}_1$ contains $T$. Then miner $M$ is incentivized to include all bounties against proofs whose Merkle trees contain $T'$. As a result, the miners of those proofs, roughly half the total in expectation, will lose their reward.

\section{Properties of the $k$-OS Criterion}
\label{sec:prop}

Recall from Section~\ref{sec:introduction} that $V_i$ represents the $i$th lowest hash value achieved after $h$ hashes are performed during interval $I$. Next, define $X_i$ to be the number of hash intervals required for the $i$th order statistic to fall below target $v$ when $h$ hashes are performed per interval. In Appendix~\ref{sec:distributions} we show that $V_i$ and $X_i$ are gamma distributed with shape parameter $i$, only differing in scale parameter. Specifically, $V_i \sim \texttt{Gamma}(t; i, v)$ and $X_i \sim \texttt{Gamma}(x; i, 1/r)$ where $v$ is the expected value of the minimum hash during interval $I$, and $r$ is the rate at which hashes are generated below $v$ during interval $I$. The two parameters are further shown to be related by the following equation
\begin{equation}
\label{eq:fund_rel}
v = r \frac{S}{h}.
\end{equation}

\subsection{Target Adjustment}

Assuming ideal difficulty adjustment, the mining target $t_k$ corresponding to mining statistic $W_k$ is set so that $t_k = E[W_k]$. In Appendix~\ref{sec:distributions}, we defined $v$ as the value of the minimum hash during interval $I$, which implies that $v = E[W_1] = t_1$. That is to say, when $k=1$, the target is equal to $v$. But this same target will produce later blocks for $k > 1$ because $W_j$ includes higher order statistics than $W_i$ when $j > i$, meaning that the network must produce more hashes in expectation in order for $W_j$ to achieve the same value as $W_i$. Therefore, the goal in this section is to determine how to select $t_k$ as a function of $v$ such that the number of hash intervals required to ensure $E[W_1] < v$ is the same as the number required to ensure $E[W_k] < t_k$. 

To that end, we begin by deriving an expression for the expected number of hash intervals required for $W_k$ to fall below target $t_k$, which we find is easiest to determine relative to a slightly different statistic. The sample average of the $X_i$ (analogous to $W_k$) is an obvious candidate, however, it actually gives the average number of intervals required for each order statistic $V_i$ to independently fall below value $v$ given that $h$ hashes are performed during each interval. Naturally, more intervals are required as $i$ increases because larger order statistics have higher expected value. So what we would like is to \emph{tune} each $X_i$ so that they are all expected to finish at the same time. 

Recall from Appendix~\ref{sec:distributions} that $E[X_i] = \frac{i}{r}$. Thus, by dividing each $X_i$ by $i$ we can align the $X_i$ to share the same expected value. To that end, define the \emph{normalized} hash interval count by $X'_i = X_i / i$; and analogously define $V'_i = V_i / i$. Each $X'_i$ is interpreted as the number of hash intervals required for $V_i$ to fall below $iv$ (or for $V'_i$ to fall below $v$). We have $E[X'_i] = \frac{1}{r}$ for all $i \geq 1$, which implies that if we tune the hash threshold to be $iv$ for $V_i$, then we expect all $V_i$ to cross below their threshold after $1/r$ hash intervals. 
\begin{mythm}\label{thm:hash_equiv}
In expectation, $1/r$ hash intervals are required to ensure that $V_i/i < v$ for all $i$.
\end{mythm}
The following result is also straightforward.
\begin{mythm}\label{thm:expected_wk}
The expected value of $W_k$ is
\[
E[W_k] =  \frac{k+1}{2} v
\]
\end{mythm}
\begin{myproof}
\begin{equation}
\label{eq:expected_wk}
\begin{array}{rcl}
E[W_k] & = & E \left[ \frac{1}{k} \sum\limits_{i=1}^k V_i \right] \\
& = & \frac{1}{k} \sum\limits_{i=1}^k E\left[ V_i \right] \\
& = & \frac{1}{k} \sum\limits_{i=1}^k iv \\
& = & \frac{k+1}{2} v. 
\end{array}
\end{equation}
\end{myproof}
And from Theorem~\ref{thm:expected_wk} it follows that
\begin{equation}
\label{eq:relate_v_vprime}
\begin{array}{rcl}
E \left[ \frac{1}{k} \sum\limits_{i=1}^k V'_i \right] & = & \frac{1}{k} \sum\limits_{i=1}^k E\left[ V'_i \right] \\
& = & v \\
& = & E[W_1] \\
& = & \frac{2}{k+1}  E[W_k]. \\
\end{array}
\end{equation}
Equation~\ref{eq:relate_v_vprime} implies that the expected sample average of values $V'_i$ is equivalent to the expected value of $W_1$, which is in turn equivalent to scaling the expected value of $W_k$ by a factor of $2 / (k+1)$.  Recalling that $t_i = E[W_i]$ when targets are tuned optimally, Theorem~\ref{thm:hash_equiv} and Equation~\ref{eq:relate_v_vprime} give us the following.
\begin{mythm}\label{thm:target}
In expectation, the same number of hash intervals are required to ensure $W_k < t_k$, for all $k > 0$, provided that $t_k$ is chosen such that
\begin{align}
t_k = \frac{k+1}{2} v.
\end{align}
\end{mythm}

\subsection{Estimating Hash Rate}
\label{sec:estimating_hash_rate}

Because $V_i$ and $X_i$ have the same distribution, up to a change of variables, Equation~\ref{eq:relate_v_vprime} also implies
\begin{equation}
\label{eq:expected_yk}
\begin{array}{rcl}
E\left[ \frac{1}{k} \sum\limits_{i=1}^k X_i \right] & = & \frac{k+1}{2} \frac{1}{r} \\
& = & \frac{k+1}{2} E\left[ \frac{1}{k} \sum\limits_{i=1}^k X'_i \right]. \\
\end{array}
\end{equation}
Therefore, assuming that targets have been adjusted so that $t_k = \frac{k+1}{2} v$, the following is a natural choice of estimator for the overall number of hash intervals required to ensure $W_k < t_k$.
\begin{equation}
Y_k =  \frac{2}{k+1} \left( \frac{1}{k} \sum_{i=1}^k X_i \right) .\label{eq:yk}
\end{equation}
Note that the estimator $Y_k$ holds the property
\[
\frac{2}{k+1} E[W_k] = E[W_1] = v = vr E[Y_1] = vr E[Y_k].
\]
In other words, $E[W_k]$ is related to $E[Y_k]$ by the constant transformation $\frac{2}{vr(k+1)}$. Equation~\ref{eq:expected_yk} shows that $Y_k$ is the sample average of random variables $X'_i$, each representing the number of hash intervals required for $V_i$ to fall below $iv$. On the other hand, Theorem~\ref{thm:hash_equiv} establishes that $E[V_i] = iv$ after mining for $1/r$ hash intervals. Therefore, $Y_k$ is biased to the extent that the sample average of the $V_i$'s deviates from its expected value. But as $k$ increases, the law of large numbers ensures that individual departures from the mean for each $V_i$ are cancelled out. Thus $Y_k$ is a consistent estimator in the limit that $k$ approaches infinity. From this argument and Equations~\ref{eq:expected_yk} and~\ref{eq:yk} we have the following. 
\begin{mythm}\label{thm:expectation_yk}
Assuming that $t_k = \frac{v(k+1)}{2}$, $Y_k$ is a consistent estimator of the expected number of hash intervals required for $W_k$ to fall below $t_k$ with
\begin{align}
E[Y_k] =  \frac{1}{r},\label{eq:expectation_yk}
\end{align}
where $r = vh / S$.
\end{mythm}

\subsection{Improvement in Variance} 

We next turn our attention to quantifying the improvement in mining time variance that is realized by using the $k$-OS criterion over the 1-OS criterion. In Section~\ref{sec:estimating_hash_rate} we established that statistic $Y_k$ is a consistent estimator of the number of hash intervals required for mining statistic $W_k$ to fall below target $t_k = \frac{k+1}{2} v$. Here we measure the change in variance of $Y_k$ as $k$ increases, while holding its expected value constant.

\begin{mythm}\label{thm:variance-blocks}
For fixed expected block discovery time, variance decreases by fraction
$\frac{8k+4}{6(k^2+k)} = O\left( \frac{1}{k} \right)$ when using mining statistic $W_k$ instead of $W_1$.
\end{mythm}

\begin{myproof}
Theorem~\ref{thm:expectation_yk} establishes that block discovery time $Y_k$ is the same in expectation for all mining statistics $W_k$ provided that $t_k =  \frac{v(k+1)}{2}$. Therefore, the ratio of variance in $Y_k$ to the variance in $Y_1$ estimates the reduction in block time variance due to \bobtail.  
\begin{equation}
\label{eq:variance-blocks}
\renewcommand*{\arraystretch}{2}
\begin{array}{rcl}
\frac{Var[Y_k]}{Var[Y_1]} & = & \frac{Var\left[ \frac{2}{k+1} \left( \frac{1}{k} \sum_{i=1}^k X_i \right) \right]}{Var[X_1]} \\
& = & \frac{4}{(k+1)^2} \frac{Var\left[ \left( \frac{1}{k} \sum_{i=1}^k X_i \right) \right]}{Var[X_1]} \\
& = & \frac{4}{(k+1)^2} \frac{(k+1)(2k+1)}{6k} \\
& = & \frac{8k+4}{6(k^2+k)} \\
& = & O\left( \frac{1}{k} \right),
\end{array}
\end{equation}
where we use the expression for $Var\left[ \left( \frac{1}{k} \sum_{i=1}^k X_i \right) \right]$ derived in Appendix~\ref{thm:variance-blocks}.
\end{myproof}

\begin{figure}[t] 
 \centerline{\includegraphics[width=\columnwidth]{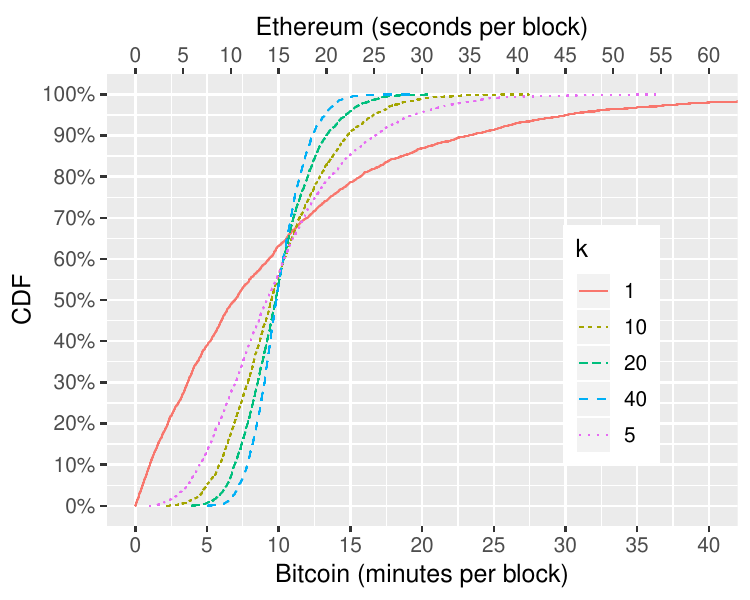}}
 \caption{Results of a Monte Carlo simulation showing the CDF of $Y_k$, the block discovery time under mining statistic $W_k$, where $k$ varies for each curve and target $t_k$ is chosen so that $E[Y_k] = E[Y_1]$. Each plot's x-axis is shown in terms of the minutes per block for Bitcoin (bottom axis) and seconds per block for Ethereum (top axis). 
 }
   \label{fig:ecdf}
\end{figure}

Figure~\ref{fig:ecdf} shows the distribution of $Y_k$ when $t_k = v(k+1)/2$ so that $E[Y_k] = E[Y_1]$. The plot shows the cumulative distribution function (CDF) based on the results of a Monte Carlo simulation\footnote{We will release all simulation source code used in this paper at camera ready.}. As the plots illustrate, the use of \bobtail mining results in a significant decrease in variance for discovering new blocks.  

\section{Network Overhead}
\label{sec:overhead}

When the mining criterion is $W_k$, $k \gg 1$, it is not efficient for each miner to send proof of work every time she finds a hash value lower than her previous best. A slight improvement to that scheme is for her to send proof of work only when her hash value is lower than the lowest $k$ hashes produced by all miners cumulatively. But even this approach will result in a large amount of network traffic early in the mining process because hash values are generated uniformly at random throughout the mining interval (see Theorem~\ref{thm:proof_share}); therefore the $k$ lowest are unlikely to be generated in a short period of time at the beginning. 

To improve network efficiency significantly we require that miners do not send or forward proof sets unless there is a high probability that the resulting proof will be part of the next block. To create such a filter, we find the largest proof value that has a probability $p$ of being part of the block given $k$; where typically, $p=0.999999$. The following theorem lets us determine the expected number of proofs forwarded per block.

\begin{mythm}\label{thm:filter}
For a given threshold probability $p$, the number of proofs announced to the network is \mbox{$y=\texttt{Quantile-Gamma}(p;k,1)$}.   
\end{mythm}

\begin{myproof} 
We know from Theorem~\ref{thm:os_cont} that the $k$th order statistic $V_k$ has distribution $\texttt{Gamma}(k,v)$. Let 
\[ G(x;k,v)=P(V_k \leq x)=p\] be the CDF of $V_k$ and $G^{-1}(p;k,v)$ be equal to $\texttt{Quantile-Gamma}(p;k,v)$, which returns $x$, the value for which $G(x;k,v) = p$. Note from Eq.~\ref{eq:fund_rel} that $v=S/h$ when $t_k$ is tuned for blocks to be generated in time $I$ (i.e. $r=1$). We expect $h$ hashes per block interval, and each has probability $x/S$ of being below the threshold. Therefore, the random variable representing the number of proofs forwarded by all miners follows distribution \texttt{Binomial}$(n=h,p=x/S)$, 
which has expectation: 
\begin{eqnarray}
\frac{h x}{S} &=& \frac{1}{v} \cdot \texttt{Quantile-Gamma}(p;k,v) \notag\\
&=& \texttt{Quantile-Gamma}(p;k,v/v)\notag\\
&=& \texttt{Quantile-Gamma}(p;k,1)
\end{eqnarray}
\end{myproof}

\noindent Notably, the value is quite low, and it is independent of $h$ the expected number of hashes required to mine a block and the size of the hash space, $S$. 
We can also use a Chernoff bound for the binomial distribution to bound the deviation in the number of messages $M$. Let $y = \frac{h x}{S}$ where $x = G^{-1}(p;k,v)$ as defined in the proof above.
\begin{eqnarray}
P(M \geq (1+\epsilon)y)\leq e^{\frac{-y\epsilon^2}{2+\epsilon}}
\end{eqnarray}
This is a tight bound, and it decreases exponentially with $y$ and similarly with $k$. For example, when $k=2$  and $p=0.999999$, then $y \approx 16.7$, and we see that $P(M>1.9y)\leq 0.0095$. For $k=3$, the probability decreases to $0.004$, and so on. 

\section{Increasing Consensus}
\label{sec:consensus}
Even when all miners operate honestly, current blockchain systems frequently suffer from orphaned blocks during their operation that diminish security and delay consensus. Orphans are  generated when the announcement of a new block by one miner takes time to propagate to all other miners. In the interim, a second miner may produce a valid block. At that point, the subset of miners who received the first block first will attempt to build upon it, and the remaining miners will build upon the second. Eventually the blockchain will fork on just one of those blocks, orphaning the other. If the set of transactions in the two blocks is not the same, then consensus is delayed. 
While the occurrence of orphans in Bitcoin is relatively low, Ethereum's use of a 15 second average block discovery time increases its orphan rate significantly. 

In Appendix~\ref{sec:distributions}, we show that $X_i$, the number of block intervals required to mine the $i$th order statistic, has distribution $\texttt{Gamma}(i, 1/r)$, where $r$ is the rate at which hashes are generated below the target. It follows that $X_1$ represents the block inter-arrival time, and it has distribution $\texttt{Exponential}(1/r) = \texttt{Exponential}(T)$, where $T$ is the expected block time. Therefore, in existing PoW blockchains, the probability that one or more other blocks will be discovered during propagation time $\tau$ is bounded by $1-\frac1{e^{\tau/T}}$. Note that this bound is pessimistic in that it assumes the worst case scenario where the author of the first block has a negligible percentage of the total network hash rate.

In this section, we examine the orphan rate associated with \bobtail mining compared to Bitcoin and Ethereum. We show that when miners follow the \bobtail protocol, orphans are strictly less likely. 

\subsection{Orphan Prevention Measures}
\label{sec:orphan_prevention}
The principal cause of orphans in \bobtail is the fact that, once more than $k$ proofs have been disseminated, there exist a combinatorial number of $k$-element subsets of those proofs. Thus at the time when there exists some subset of $k$ proofs whose mean falls below target $t$, there is a reasonable chance that \emph{some other} subset also exists (or will exist relatively soon). Fortunately, the proof package rules introduced in Section~\ref{sec:protocol} greatly reduce the number of valid subsets. First, all proofs must be tethered to a \emph{supporting proof}, the latter of which should be the smallest proof previously received by the miner. Second, no support in the proof package can have value less than the 1OS. And third, the block must be signed by the private key used to generate the 1OS. 

Together, these conditions ensure that if at least one of the $k$ proof sets in the proof package points to the 1OS as support (excluding the 1OS itself), then the creator of the 1OS is the \emph{only} miner capable of assembling that proof package. Conversely, although another miner, say the one who generated $V_2$, might be capable of collecting a set of proofs that exclude the 1OS but still having mean below target $t$, that miner cannot assemble a proof package if even a single proof set includes the 1OS as support.  

\begin{figure}[t]
   \centering
   \includegraphics[width=.9\columnwidth]{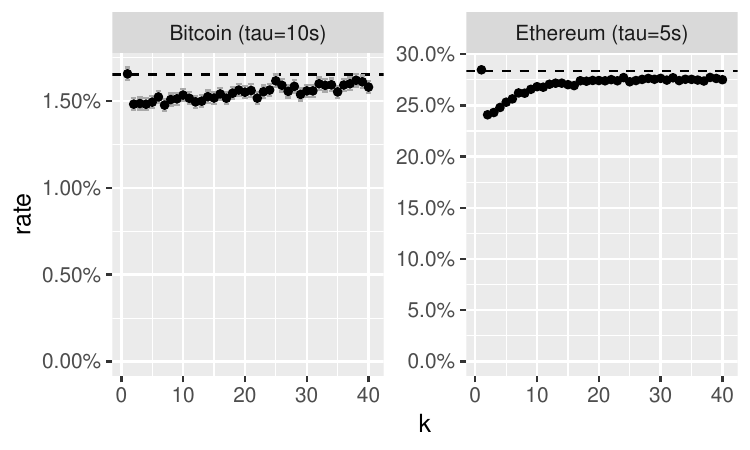} 
   \caption{A simulation of \bobtail's orphan rate when proofs and blocks propagate with constant delay of $\tau=10$ seconds and the interblock time is $T=600$ seconds. The orphan rate of \bobtail is at or below the expected orphan rate, $1-1/e^{\tau/T}$ for $k=1$, which is shown as a dashed line. Similar results hold for Ethereum where $\tau=5$ and $T=15$ seconds. Error bars represent a 95\% c.i.}
   \label{fig:orphans}
\end{figure}

\subsection{Performance}

We ran a discrete event simulator to determine the efficacy of the orphan prevention measures described in Section~\ref{sec:orphan_prevention}. The simulation includes only honest miners: once it has received a valid proof package, an honest miner does not release a competing proof package of its own. We evaluate attacks on \bobtail subsequently.  

The simulation generates blocks by repeatedly selecting values uniformly at random between $0$ and $2^{32}$. The smallest $k$ values are used as a candidate block given a pre-set target value. The propagation delay of new proofs and blocks is $\tau$ seconds. Once a block is found, we assume that the authoring miner drops out and her mining power is replaced by a new honest miner; i.e., the hash rate does not change. For $\tau$ seconds, the miners continue seeking a new block following the rules in Section~\ref{sec:protocol}. For example, if they find a block is possible with a higher 1OS, they will not release the new block. 

Figure~\ref{fig:orphans} (left) shows the results for a Bitcoin-like scenario where the inter-block time is targeted at $T=600$ seconds and the propagation delay is $t=10$ seconds. The orphan rate for $k=1$ follows the expected Poisson result, shown as a dashed line. The experiment shows that \bobtail has an orphan rate at or below the $k=1$ rate, in terms of statistical significance. 
Figure~\ref{fig:orphans} (right) shows the same  result for a simulation of Ethereum where $\tau=5$ and $T=15$ seconds, respectively.

\section{Incentivizing Honest Behavior with Rewards}
\label{sec:rewards}
In this section, we show that there exists a \emph{reward scheme} --- the payout of fees and coinbase --- that incentivizes miners to \1 continue mining for increased reward (rather than stopping once any proof is discovered) \2 use the lowest  proof they know of as support, and \3 immediately broadcast all sufficiently low proofs. 
In this section, we evaluate the following reward structure with respect to all three properties assuming honest miner behavior (attack scenarios are considered in Section~\ref{sec:dbl_self}).
\begin{itemize}
\item To each proof in the proof package, we assign \emph{primary reward} $R$, which is the same for each proof in a given package, but may vary from block to block.
\item To each proof whose support is the 1OS, we award a \emph{bonus reward} $B$, which is again the same for every proof pointing to the 1OS, but may vary by block.
\end{itemize}
One of the major goals of this section is to determine the expected primary and bonus rewards accrued by an honest miner across all proofs in a given block. We further derive the expected \emph{total reward} $T$, which is the sum of expected primary and bonus rewards for a miner following the honest strategy.

\subsection{Idealized Analysis}
\label{sec:ideal_analysis}

We begin with a basic result that is useful in contemplating reward distribution.

\begin{mythm}
\label{thm:proof_share}
In expectation, a fraction $x$ of the mining power will generate a fraction $x$ of all proofs as well as a fraction $x$ of the $k$-lowest order statistics.
\end{mythm}

\begin{myproof}
Without loss of generality, assume a single miner $M$ owns fraction $x$ of the mining power. All hashes generated are uniformly distributed throughout space $S$. Therefore, of all the hashes that fall within an interval of $S$, miner $M$ will own fraction $x$. The interval $[0, S]$ contains all proofs; it is therefore clear that $M$ will generate fraction $x$ of all proofs. Moreover, the set of all proofs $K$ that are at or below the $k$th order statistic defines an interval, $[0, k\text{-OS}]$. Thus $M$ will own fraction $x$ of proofs in $K$ as well, which constitutes fraction $x$ of the set of lowest $k$ order statistics.
\end{myproof}

We next analyze the reward payout with respect to our desired mining properties under the assumption that all miners behave honestly, i.e.\ according to the protocol. Consider a miner $M$ who possesses fraction $x$ of the total mining power. According to Theorem~\ref{thm:proof_share}, $M$ can expect to have generated fraction $x$ of the $k$ proofs in the proof package. Therefore, $M$ will earn $xkR$ primary reward in expectation. The expected bonus reward is straightforward to calculate as well, but requires the following observation. 

\begin{mythm}
\label{thm:value_corr}
The rank (i.e., hash value) of a proof is uncorrelated with the time it is generated.
\end{mythm}

\begin{myproof}
Let $T = P_1, P_2, \ldots$ be the set of all proofs generated during time interval $I$, and assume without loss of generality that the proofs are ordered chronologically so that $P_i$ was generated before $P_j$ if $i < j$. Define $V(P)$ as the hash value of proof $P$ and let $V(T)$ denote the set of all proof values generated during $I$. It will suffice to show that the conditional probability that $P_i$ achieves a given value $v \in V(T)$ is uniform for all $P_i$. 

Being drawn from a uniform distribution, we have that $P(V(P_i) = v)$ is equal for all proofs $P_i$. Next define $V_v(T) = V(T) \setminus \{v\}$, which implies $P(V(T) ~|~ V(P_i)  = v) = P(V_v(T))$ because $P(V(P_i) )$ and $P(V(P_j))$ are independent for $i \neq j$. Thus
\[
\renewcommand*{\arraystretch}{1.5}
\begin{array}{r}
P(V(P_i) = v ~|~ V(T)) = \\
\frac{P(V(T) ~|~ V(P_i) = v)P(V(P_i) = v)}{P(V(T))} = \\
P(V(P_i) = v) \frac{P(V_v(T))}{P(V(T))} = c
\end{array}
\]
for some constant c.
\end{myproof}

We can use Theorem~\ref{thm:value_corr} to show that half of a miner's proofs in the proof package are expected to be generated after the 1OS. Thus honest miner $M$, with fraction $x$ of the hash rate, will generate $\frac{xk}{2}$ proofs that use the 1OS as support. It follows that $M$'s expected bonus reward is equal to $\frac{xkB}{2}$. Finally, the expected total reward for the honest miner is given by 
\begin{equation}
T_H = xk \left( R + \frac{B}{2} \right).
\label{eqn:hon_tot_rew}
\end{equation}

From this expression for total reward, we can see that honest mining delivers all three desired mining properties. First, a miner's reward is proportional to her hash rate, which encourages her to mine as much as possible rather than stopping once a proof is found. Second, her total reward is an increasing function of the number of her proofs that point to the 1OS. And third, because total reward is also an increasing function of the number of proofs in the proof package, she is incentivized to release her proofs as soon as possible so as to give them greatest chance of being included.

\begin{figure}[t] 
   \centerline{\includegraphics[width=0.8\columnwidth]{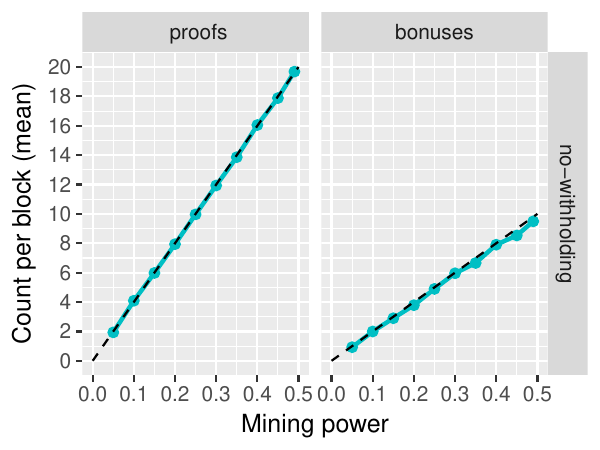}}
   \caption{Everyone is honest. The dotted lines show the  predicted value of $R$ and $B$ from Eq.~\ref{eqn:hon_tot_rew}.}
   \label{fig:no-withholding}
\end{figure}

Figure~\ref{fig:no-withholding} shows the results of this rewards scheme from a simulation of honest miners. The dotted lines show the values predicted by Eq.~\ref{eqn:hon_tot_rew}. In the next section, we demonstrate that dishonest miners earn only fewer rewards. 

\section{Thwarting Attacks}
\label{sec:dbl_self}

We next demonstrate quantitatively that its reduced inter-block-time variance allows Bobtail to thwart both {\em doublespend}~\cite{Nakamoto:2009} and {\em selfish mining}~\cite{Eyal:2014} attacks. We further show that while Bobtail  introduces the possibility of a new {\em proof withholding} attack, a simple protocol policy ensures that attackers receive substantially lower reward when carrying out this attack.

\subsection{Doublespend and Selfish Mining Attacks}

\begin{figure}[t]
   \centering
   \includegraphics[width=\columnwidth]{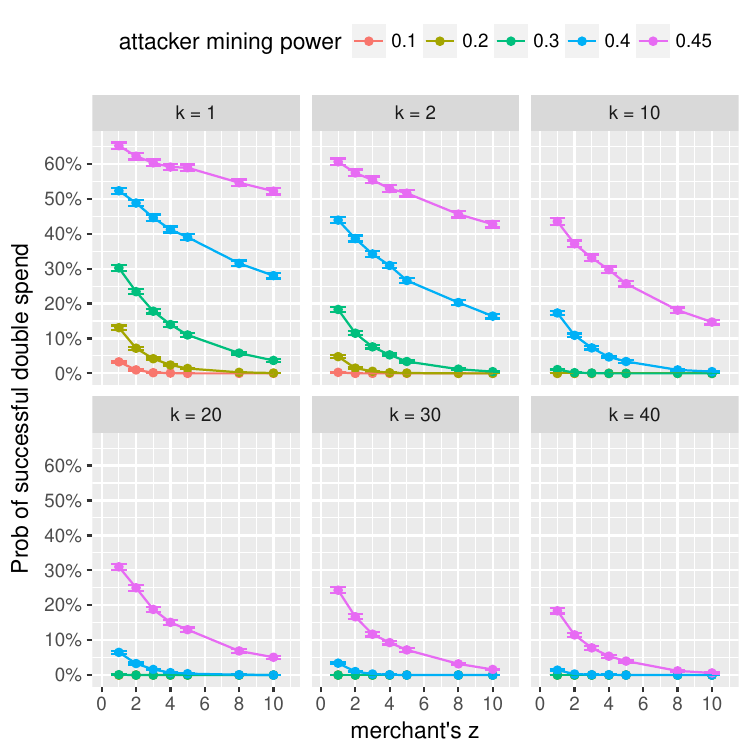} 
   \caption{Doublespend attack success given $k$ for various values of attacker mining 
power (each line) and merchants embargo period $z$ (on the independent axis). Error 
bars show 95\% c.i.'s.  }
   \label{fig:dblspend}
\end{figure}

Figure~\ref{fig:dblspend} shows a Monte Carlo simulation of the doublespend attack. The 
merchant has setup an embargo period of $z$ blocks.  The attacker's strategy is to mine 
until its branch is longer than that of the honest miners, or until the honest branch is ahead 
by $3z+5$ (to ensure that the attack has finite duration). Each facet of the plot represents a value of $k$. The results show that as 
$k$ increases and variance decreases, the probability of attacker success significantly 
decreases. For example, in today's implementations of both Bitcoin and Ethereum, an attacker with 40\% of the mining power will succeed with 30\% probability when $z=8$; however, using Bobtail with $k\geq20$, the probability of success falls to less than 1\%. 

\begin{figure}[t]
   \centering
   \includegraphics[width=.9\columnwidth]{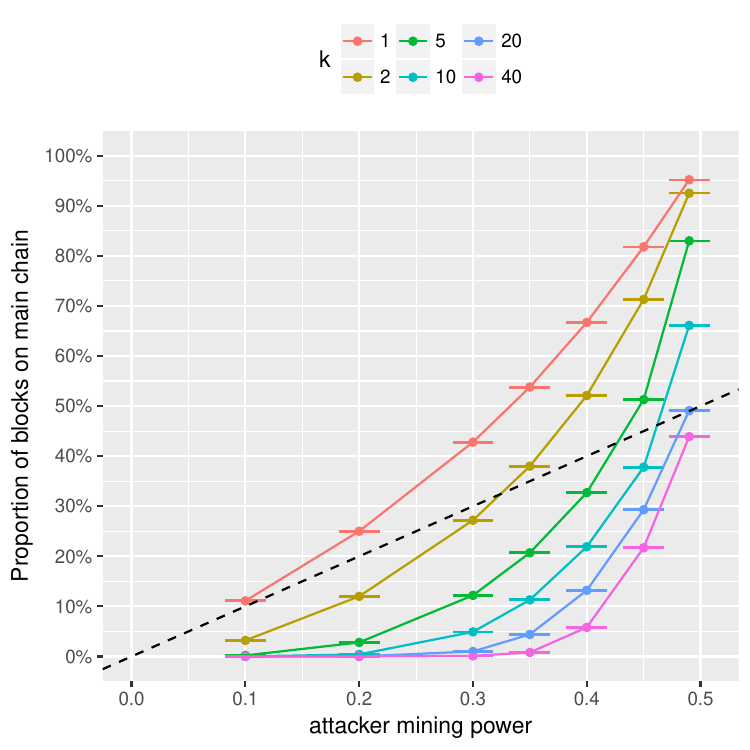} 
   \caption{Selfish mining attack success given $k$ (different lines) for various 
values of attacker mining power (on the independent axis).  The dashed line shows the 
results of honest mining. Error bars show 95\% c.i.'s. }
   \label{fig:selfish}
\end{figure}

Figure~\ref{fig:selfish} shows a similar result for selfish mining via a Monte Carlo 
simulation. The attacker follows the selfish mining strategy and it is assumed that, during a block race, the attacker's block always propagates to miners before the block of any other honest miner. The 
figure shows the proportion of blocks on the main chain won by attackers. The dashed 
line represents the proportion that would be won by honest mining. For example, a 
selfish miner with 40\% of the mining power will claim about 66\% of blocks with 
Bitcoin and Ethereum currently; however, using Bobtail with $k\geq5$, the same miner will find that 
selfish mining is less successful than honest mining. 

\subsection{Zero-cost Zero-confirmation (ZCZC) Doublespend Attacks}

Our approach to mitigating ZCZC doublespends is to simply withhold reward from the offending party. Suppose that proof set $\mathcal{P}_i$ is associated with transaction set $\mathcal{T}_i$, which contains $T'$. Suppose further that $\mathcal{T}_1$ contains $T$, a transaction incompatible with $T'$. As stipulated in Section~\ref{sec:protocol}, the miner assembling the block can include bounty $\mathcal{B}(\mathcal{T}_i, T')$. We now further stipulate that both the primary and bonus rewards nominally owed to $\mathcal{P}_i$ for this block shall instead be awarded to $\mathcal{P}_1$. The implication of this policy is that it is the miner of $\mathcal{P}_1$ who determines what is the \emph{correct} transaction when there exists an incompatibility. Note that, based on this approach, it becomes possible for the miner of $\mathcal{P}_1$ to intentionally mine transactions incompatible with the other miners in order to claim their reward. However, this strategy could only be profitable in the long-run if the miner possesses more than 50\% of the hash rate. Otherwise, the miner's proofs will most often be something other than $\mathcal{P}_1$, in which case they will be the ones to lose reward.

\subsection{Proof Withholding Attacks}

Bobtail allows for an attack where a malicious miner withholds proofs for a competitive advantage. In this section, we demonstrate that our design of Bobtail ensures the economic rewards for withholding attackers is substantially lower than that of honest miners.

In the withholding attack, the malicious miner does not announce her own proofs to the other miners. This behavior can be advantageous in two ways. First, it gives her more time to become the 1OS, which means she controls the set of transactions included in the block, $\mathcal{T}_1$. Second, it allows the attacker to \emph{pack} more of her own proofs into the proof package if she does manage to mine the 1OS. The attacker mines until either she is able to assemble a block as author of the 1OS; or it is clear that the honest miners are more likely to release a block without her withheld proofs. In the latter case, she disseminates her withheld proofs, hoping that some will be included in the proof package of the next block. 

To thwart the attack, Bobtail includes two simple rules. First, it is considered honest behavior for miners to prioritize inclusion of their own proofs when assembling a block. Second, after prioritizing their own proofs, if multiple subsets of $k$ proofs can be used to assemble a block, an honest miner will select proofs from other miners in the order that they were received locally over the network. In other words, the proofs of the withholding attacker will be likely left out if withheld too long.
 
\begin{figure}[t] 
   \centerline{\includegraphics[width=0.8\columnwidth]{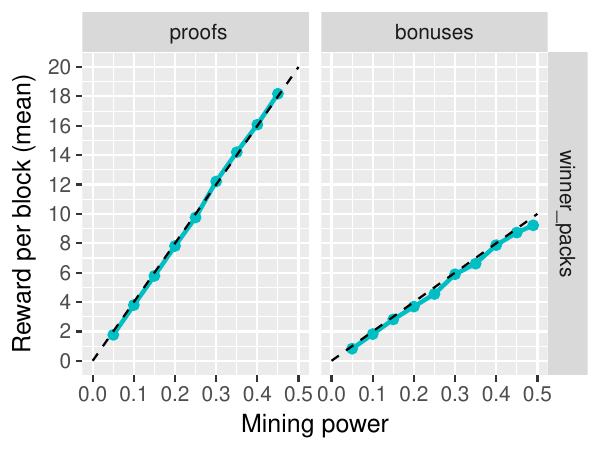}}
   \caption{All miners behave honestly. The dashed lines show the  predicted value of $R$ and $B$ from Eq.~\ref{eqn:hon_tot_rew}.}
   \label{fig:no-withholding-winner-packs}
\end{figure}

\begin{figure}[t] 
   \centerline{\includegraphics[width=0.8\columnwidth]{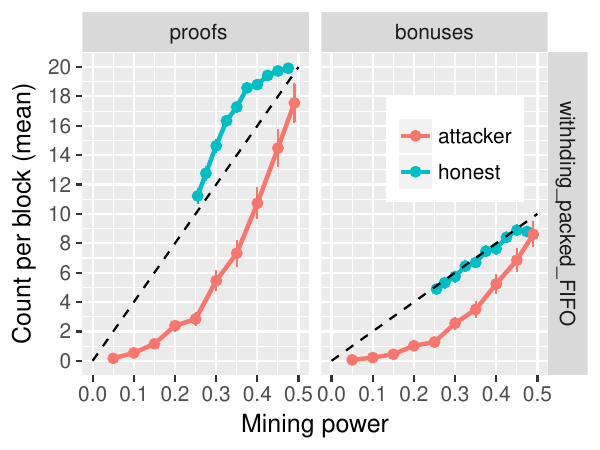}}
   \caption{Withholding by attackers results in greater rewards for the honest miners.  The dashed lines show the  predicted value of $R$ and $B$ from Eq.~\ref{eqn:hon_tot_rew}. }
   \label{fig:withholding-fifo}
\end{figure}

We evaluated this attack using a Monte Carlo simulation. Figure~\ref{fig:no-withholding-winner-packs} shows the allocation of rewards and bonuses when all miners act honestly: they are precisely predicted by Eq.~\ref{eqn:hon_tot_rew}. Figure~\ref{fig:withholding-fifo} shows result of the withholding attack given the two rules described above. Malicious behavior results in the attacking miner receiving lower reward from proofs and bonuses than honest miners. This is because the profit that she loses for releasing her proofs too late when she does not mine the block is greater than the profit she gains by withholding when she does.  Furthermore, the rewards to honest miners are actually greater because of the attack.

\subsection{Denial-of-Reward Attacks}

In contrast to doublespend and selfish mining attacks, a denial-of-reward (DoR) attack can be carried out by any network participant, not just a miner. The attacker releases two incompatible transactions $T$ and $T'$, each to disjoint subsets of the miners. The result is that any proof $\mathcal{P}_i$ generated such that $T \in \mathcal{T}_i$, will receive no reward if $\mathcal{P}_1$ is generated such that $T' \in \mathcal{T}_1$. In this way, the attacker can lower the profitability of mining for all or a subset of miners.

When network latency between miners is reasonably low, and assuming that an eclipse attack~\cite{Heilman:2015} on miners is not possible, DoR attacks can be rendered largely ineffective. Even if half the miners initially receive transaction $T$, while the other half receive incompatible transaction $T'$, all miners will receive both $T$ and $T'$ within seconds. In Bitcoin, the probability that one of the $k$ lowest OSes is mined within a time-period of a few seconds is very low. In Ethereum, the probability would be much higher, but miners could simply adopt a policy of waiting several seconds before beginning to mine a newly received transaction. 

With knowledge of the existence of $T$ and $T'$, the safest strategy for miners is to exclude both. However, this practice can leave an honest transaction creator stuck (who might have accidentally submitted two transactions spending the same UTXO) and also prevents miners from collecting the associated fee. Thus, the best approach is for miners to establish the following convention. If $T'$ is received more than a few seconds after $T$, then discard $T'$ and mine $T$ exclusively. When $T$ and $T'$ are received within a few seconds of each other, mine the transaction with the lowest hash value if the fees are the same, otherwise mine the transaction with the highest fee. 
Note that miners have ample incentive to follow this convention because if they do not, then there is a good chance that the proofs they mine will be incompatible with the 1OS, and therefore will receive no reward. 

\section{Conclusion}

We have designed and characterized a novel method of low-variance blockchain mining. 
We have derived expressions for the expectation and variance of the \bobtail mining proof and the number of hashes performed for any value of $k$. Using these expressions, we have shown that \bobtail reduces variance by a factor of $O(1/k)$, compared to using $k=1$. 
 We have also shown that forks are created by \bobtail miners no more often than existing systems, and that  dishonest  miners  receive significantly lower rewards due to minor protocol adjustments.  Furthermore, we have demonstrated that low-variance mining significantly reduces the effectiveness of doublespend and selfish mining attacks, and that our design thwarts withholding and denial-of-reward attacks. Finally we have introduced a policy for proof dissemination that keeps network traffic to a minimum.

\bibliographystyle{IEEEtran}
\bibliography{references}

\appendix

\section{Distributions of mining processes}
\label{sec:distributions}

Consider the distribution of $H$ an arbitrary random variable chosen from the sequence of block hashes $H_1, \ldots, H_h$. We have $f_{H}(t;S) = 1/S$ and $F_{H}(t;S) = t/S$. The following result is well known\footnote{See for example, Casella and Berger~\cite{Casella:2002}}.

\begin{mylem} \label{lem:pdf}
The probability density function (pdf) of the $i${\rm th} order statistic, $V_i$, from $h$ samples (i.e., hashes) is 
\begin{equation} 
\renewcommand*{\arraystretch}{2}
\begin{array}{rcl}
f_{V_i}(t; S,h) & = & \frac{h!}{(i-1)! (h-i)!} f_{H}(t) \left( F_{H}(t) 
\right)^{i-1} \left(1 - F_{H}(t) \right)^{h-i} \\
 & = & \frac{h!}{(i-1)! (h-i)!} \frac{1}{S} \left( \frac{t}{S} 
\right)^{i-1} \left(1 - \frac{t}{S} \right)^{h-i}.
\end{array}
\label{eq:pdf_os}
\end{equation}
\end{mylem}

\noindent When hash interval $I$ corresponds to the desired block time, say 10 minutes for Bitcoin, there will be many hashes performed during the interval. So it is reasonable to consider how the distribution for $V_i$ changes in the limit that $h$ approaches infinity. 

\begin{mythm}
\label{thm:os_cont}
In the limit that $h$ approaches infinity, $V_i \sim \texttt{Gamma}(i, v)$ where $v$ is the expected value of the minimum hash.
\end{mythm}\smallskip

\begin{myproof}
Define $g(t; i, v)$ to be the PDF of the Gamma distribution with shape parameter $i$ and scale parameter $v$. If the number of hashes approaches infinity, then so must the size of the hash space, and yet $S$ must always be larger than $h$. Therefore, we assume that $h = S / v$ for arbitrary parameter $v > 1$. Under this assumption we can equivalently consider limit that $S$ approaches infinity. We have

\begin{align}
\renewcommand*{\arraystretch}{2}
\begin{array}{rcl}
\label{eqn:univariate_}
f_{V_i}(t; S, h)  & =& \lim\limits_{h \rightarrow \infty} f_{V_i}(t; S, h) \\ 
& =& \resizebox{.7\hsize}{!}{$\lim\limits_{S \rightarrow \infty} \frac{(S/v)!}{(i-1)! 
\left( \frac{S}{v}-i \right)!} \frac{1}{S} \left( \frac{t}{S} \right)^{i-1} \left(1 - 
\frac{t}{S} \right)^{\frac{S}{v}-i}$}\\
& = & \lim\limits_{S \rightarrow \infty} \frac{(S/v)!}{S^i  (i-1)! \left( \frac{S}
{v}-i \right)!} t^{i-1} \left(1 - \frac{t}{S} \right)^{\frac{S}{v}-i} \\
&=& \resizebox{.7\hsize}{!}{$\frac{t^{i-1}}{(i-1)!} \left[ \lim\limits_{S \rightarrow 
\infty} \frac{\left( \frac{S}{v} \right) \ldots \left( \frac{S}{v}-i+1 \right)}{S^i} 
\right] \left[ \lim\limits_{S \rightarrow \infty} \left(1 - \frac{t}{S} \right)^{\frac{S}
{v}-i} \right]$} \\
& = & \frac{t^{i-1}}{(i-1)! v^i} e^{\frac{-t}{v}} \\
& = & g(t; i, v),
\end{array}
\end{align}
The second to last step follows from the fact that
\begin{equation}
\lim\limits_{S \rightarrow \infty} \frac{\left( \frac{S}{v} \right) \ldots 
\left( \frac{S}{v}-i+1 \right)}{S^i} = \lim\limits_{S \rightarrow \infty} 
\frac{ \left( \frac{S}{v} \right)^i }{S^i} = \frac{1}{v^i},
\end{equation}
and the common limit
\begin{equation}
\lim\limits_{S \rightarrow \infty} \left(1 - \frac{t}{S} \right)^{\frac{S}{v}} = 
e^{\frac{-v}{v}},
\end{equation}
which implies that
\begin{align}
\!\!\!\lim\limits_{S \rightarrow \infty} \left(1 - \frac{t}{S} \right)^{\frac{S}{v}-i}& 
= 
 \left[ \lim\limits_{S \rightarrow \infty} \left(1 - \frac{t}{S} \right)^{-i} \right] 
\left[  \lim\limits_{S \rightarrow \infty} \left(1 - \frac{t}{S} \right)^{\frac{S}{v}} 
\right] \\
& =  1 \cdot e^{\frac{-t}{v}}.
\end{align}
When $i=1$, $V_1 \sim \texttt{Gamma}(t; 1, v) = \texttt{Exponential}(t; v)$. And since the expected value of an exponential random variable is equal to the value of its scale parameter, we can see that $v$ is simply the expected value of the minimum hash.
\end{myproof}

Next, consider the PDF of $X_i$, $f_{X_i}(x; S, v)$. After $x$ hash intervals, let $E$, $L$, and $G$ be, respectively, the events that the $i$th order statistic is equal to $v$, the order statistics below $i$ are less than $v$, and the order statistics above $i$ are greater than $v$. Furthermore, let $\mathcal{O}$ be the set of all divisions of $H_1, \ldots, H_h$ into distinct sets $\{H ~|~ H = V_i\}$, $\{H ~|~ H < V_i\}$, and $\{H ~|~ H > V_i\}$. We have
\begin{equation}
\renewcommand*{\arraystretch}{2}
\begin{array}{rcl}
f_{X_i}(x; S, v) &=& \sum\limits_{o \in \mathcal{O}} P[E(x), L(x), G(x) ~|~ o] \\
f_{X_i}(x; S, v) &=& \binom{h x}{i-1}(hx-i+1) P[E(x)~|~o] P[L(x)~|~o] P[G(x) ~|~ o] \\
f_{X_i}(x; S, v) &=& \frac{(hx)!}{(i-1)!(hx-i)!} P[E(x)~|~o] P[L(x)~|~o] P[G(x) ~|~ o] \\
f_{X_i}(x; S, v) &=& \frac{(hx)!}{(i-1)!(hx-i)!} \frac{1}{S} \left(\frac{v}{S}\right)^{i-1} \left(1 - \frac{v}{S}\right)^{hx-i} \\
\end{array}
\end{equation}

Assuming that $I$ is large, it again makes sense to consider the limit as $h$ approaches infinity. 
\begin{mythm}
\label{thm:hash_cont}
In the limit that $h$ approaches infinity, $X_i \sim \texttt{Gamma}(i, 1/r)$ where $r$ is the expected number of hashes falling below $v$ during a given interval.
\end{mythm}\smallskip

\begin{myproof}
The probability that any given hash \emph{succeeds}, i.e. falls below $v$, is given by $p = \frac{v}{S}$. Again, we would like to consider the limit as $h$ approaches infinity. But in doing so, we must ensure that the probability of hash success remains constant. In other words, the probability of hash success must diminish as $h$ increases. So there must exist some constant $r$ such that $\frac{r}{h} = p = \frac{v}{S}$. It follows that
\begin{equation}
\renewcommand*{\arraystretch}{2}
\begin{array}{rcl}
f_{X_i}(x; S, v) &=& \frac{(hx)!}{(i-1)!(hx-i)!} \frac{r}{h} \left(\frac{r}{h}\right)^{i-1} \left(1 - \frac{r}{h}\right)^{hx-i} \\
\end{array}
\end{equation}
Arguing in similar fashion as for $V_i$, we find that 
\[
\lim\limits_{h \rightarrow \infty} f_{X_i}(x; S, v) = g(x; i, 1/r).
\]
Thus, $E[X_i] = 1/r$, which implies that $r$ should be interpreted as the expected rate at which hashes fall below $v$ during a single interval I. 
\end{myproof}

We can see that the distributions for $V_i$ and $X_i$ are related through the change of variables $v = 1/r$, and all four parameters $v$, $r$, $h$, and $S$ are related by
\[
v = r \frac{S}{h}.
\]
In words, the latter states: the expected value of the minimum hash is related to the expected number of blocks mined per interval by the ratio $\frac{S}{h}$. 

\subsection{Joint Distributions}

\begin{mythm}
\label{thm:joint_V}
The joint distribution of the $i$th and $j$th order statistic of uniform random samples $H_1, \ldots, H_h$  is given by
\begin{align}
f_{V_i, V_j}(t_i,t_j; v) & =  
g(t_i; i, v) g(t_j-t_i; j-i, v).\label{eqn:joint_V}
\end{align}
where $v$ is the expected value of the minimum hash.
\end{mythm}

\begin{myproof}
It is well known\footnote{See Casella and Berger~\cite{Casella:2002}, Theorem 5.4.6.} that 
the joint distribution of the $i$th and $j$th order statistics, out of $h$ total samples, is given by
\begin{align}
f_{V_i, V_j}(t_i,t_j; v) &= \resizebox{.7\hsize}{!}{$\frac{h!}{(i-1)! (j-1-i)! (h-j)!} f_H(t_i) 
f_H(t_j) [F_H(t_i)]^{i-1}$} \nonumber\\
 & \times [F_H(t_j) - F_H(t_i)]^{j-1-i} [1 - F_H(t_j)]^{n-j}.
\end{align}
Thus, we have
\begin{equation}
\renewcommand*{\arraystretch}{2}
\begin{array}{rcl}
f_{V_i, V_j}(t_i,t_j; S, v)\!\!\!& = & \frac{t_i^{i-1} (t_j-t_i)^{j-1-i}}{S^j} \frac{h!}
{(i-1)! (j-1-i)! (h-j)!} \left(1 - \frac{t_j}{S} \right)^{h-j} \\
& = & \frac{\frac{S}{v} \ldots \left(\frac{S}{v}-j+1\right)}{S^j} \frac{t_i^{i-1} (t_j-
t_i)^{j-1-i}}{(i-1)! (j-1-i)!} \left(1 - \frac{t_j}{S} \right)^{\frac{S}{v}-j}.
\end{array}
\end{equation}
Finally, assuming $j > i$, and reasoning in the limit as $S \rightarrow \infty$ in the same manner as above,
\begin{align}
\renewcommand*{\arraystretch}{2}
\begin{array}{rcl}
f_{V_i, V_j}(t_i,t_j; v) & = \lim_{S \rightarrow \infty} f_{V_i, V_j}(t_i,t_j; S, v) \\ 
&= \frac{1}{v^j} \frac{t_i^{i-1} (t_j-t_i)^{j-1-i}}{(i-1)! (j-1-i)!} e^{-\frac{t_j}{v}} \\
&= \frac{t_i^{i-1}}{v^i (i-1)!} e^{-\frac{t_i}{v}} \frac{(t_j-t_i)^{j-1-i}}{v^{j-i} 
(j-1-i)!} e^{-\frac{t_j-t_i}{v}} \\
&= g(t_i; i, v) g(t_j-t_i; j-i, v).
\end{array}
\end{align}
\end{myproof}

Because $X_i$ shares the same distribution as $V_i$, up to the change of variables $v = 1/r$, the following result follows trivially.

\begin{mythm}
The joint distribution of $X_i$ and $X_j$, $j > i$, is given by
\begin{align}
f_{X_i, X_j}(t_i,t_j; 1/r) & =  
g(t_i; i, 1/r) g(t_j-t_i; j-i, 1/r).\label{eqn:joint_X}
\end{align}
where $r$ is the expected rate at which hashes fall below $v$ during a single interval I.
\end{mythm}

\section{Moments of $W_k$}
\label{sec:limits}

The goal of this section is to empirically validate our expression for $E[W_k]$ from Section~\ref{sec:prop} and then derive and validate an expression for $Var[W_k]$. $W_k$ is simply the sample mean over the lowest $k$ order statistics $V_1, \ldots, V_k$. But, unfortunately, the analysis below is not straightforward because  the $V_i$ are neither independent nor identically distributed. 

\begin{figure}[t] 
   \centerline{\includegraphics[width=\columnwidth]{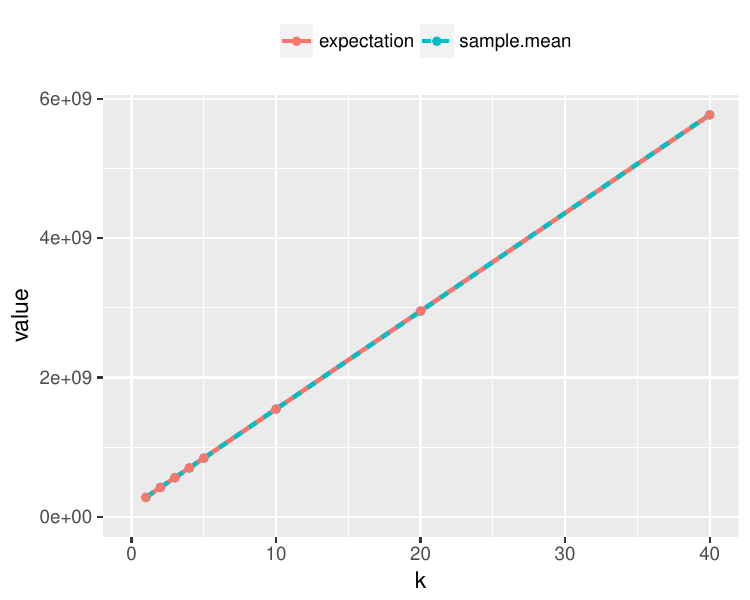}}
   \caption{Eq.~\ref{eq:expected_wk} versus simulation.}
   \label{fig:expectation}
\end{figure}
\para{Empirical Validation of Theorem~\ref{thm:expected_wk}.} Figure~\ref{fig:expectation} 
compares  Eq.~\ref{eq:expected_wk} versus a result obtained through a small Monte Carlo 
simulation of \bobtail mining run tens of thousands of times, with $k$ as the independent 
variable. In all cases, the results match perfectly. 

\subsection{Variance of $W_k$}
\label{sec:variance}

\begin{mythm}
\label{thm:variance_wk}
The variance of $W_k$ is
\begin{equation}
\label{eq:variance_wk}
Var[W_k] = \frac{(k+1)(2k + 1)}{6k} v^2. 
\end{equation}
\end{mythm}

\begin{figure}[t] 
   \centerline{\includegraphics[width=\columnwidth]{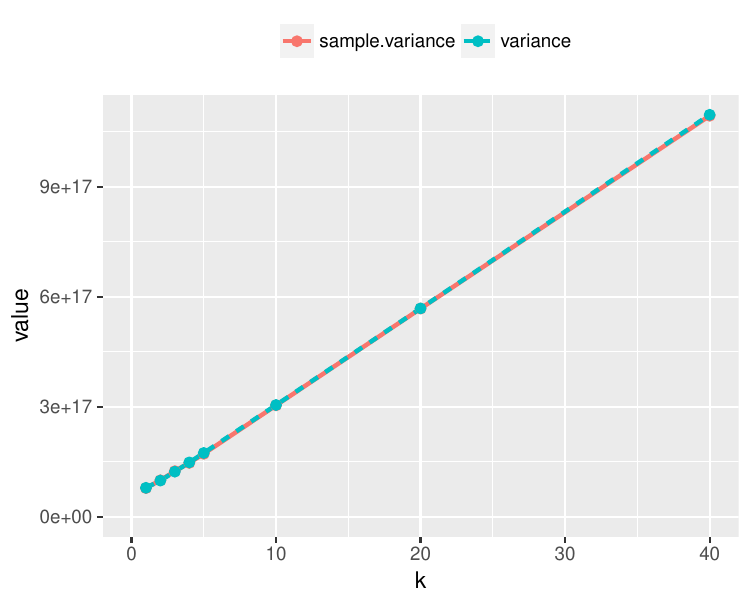}}
   \caption{Eq.~\ref{eq:variance_wk} versus simulation.}
   \label{fig:variance}
\end{figure}

\begin{myproof}
Assuming that $j > i$, Theorem~\ref{thm:joint_V} yields

\begin{equation}
\renewcommand*{\arraystretch}{2}
\begin{array}{rcl}
\label{eqn:expect_product}
E[V_i V_j] &=& \int_0^{\infty} \int_{t_i}^{\infty} {t_i} t_j g(t_i; i, v) g(t_j-t_i; j-i, v) dt_j dt_i \\
&=& \int_0^{\infty} t_i g(t_i; i, v) \int_0^{\infty} (w+t_i) g(w; j-i, v) dw dt_i \\
&=& \int_0^{\infty} t_i g(t_i; i, v) [(j-i) v + t_i] dt_i \\
&=& v (j-i) \int_0^{\infty} t_i g(t_i; i, v) dt_i + \int_0^{\infty} t_i^2 g(t_i; i, v) dt_i \\
&=& i v^2 (j-i) + i v^2(1 + i) \\
&=& i v^2 (1+j).\vspace{-1.5em}
\end{array}\vspace{1em}
\end{equation}
Before continuing, we note that since $V_i \sim \texttt{Gamma}(i, v)$, it follows
that
$Var[V_i] = i v^2$.  Now, assuming that $j > i$, and using Eq.~\ref{eqn:expect_product}, we have
\begin{equation}
\renewcommand*{\arraystretch}{2}
\begin{array}{rcl}
\label{eqn:covariance}
\texttt{cov}[V_i, V_j] &=& E[V_i V_j] - E[V_i] E[V_j] \\
&=& i v^2 (1+j) -  (i v)(j v) \\
&=& i v^2 \\
&=& Var[V_i].
\end{array}
\end{equation}
Finally, we find the variance of $W_k$ by substituting first Eq.~\ref{eq:wk} and then Eq.~\ref{eqn:covariance}:
\begin{align}
\renewcommand*{\arraystretch}{2}
\begin{array}{rcl}
Var[W_k]  &=& Var \left[\frac{1}{k} \sum\limits_{i=1}^k V_i \right] \\
 &=& \frac{1}{k^2} Var \left[ \sum\limits_{i=1}^k V_i \right] \\
&=& \frac{1}{k^2} \left( \sum\limits_{i=1}^k Var[V_i] + 2 \sum\limits_{j=1}^k \sum\limits_{i=1}^{j-1} \texttt{cov}[V_i, V_j]\right) \\
&=& \frac{1}{k^2} \left( \sum\limits_{i=1}^k i v^2 + 2 \sum\limits_{j=1}^k \sum\limits_{i=1}^{j-1} i v^2 \right) \\
&=& \frac{v^2}{k^2} \left( \frac{k(k+1)}{2}+ \sum\limits_{j=1}^k j(j-1) \right) \\
&=& \frac{v^2}{k^2} \left( \frac{k(k+1)}{2} + \frac{k(k+1)(2k + 1)}{6} - \frac{k(k+1)}{2} \right) \\
&=& \frac{(k+1)(2k + 1)}{6k} v^2. \vspace{-1em}
\end{array}
\end{align}
\end{myproof}
\para{Empirical Validation of Theorem~\ref{thm:variance_wk}.}
Figure~\ref{fig:variance} shows Eq.~\ref{eq:variance_wk} versus  our  Monte Carlo simulation where $k$ is the independent variable. The results show an exact match.

\begin{figure}[t] 
   \centerline{\includegraphics[width=\columnwidth]{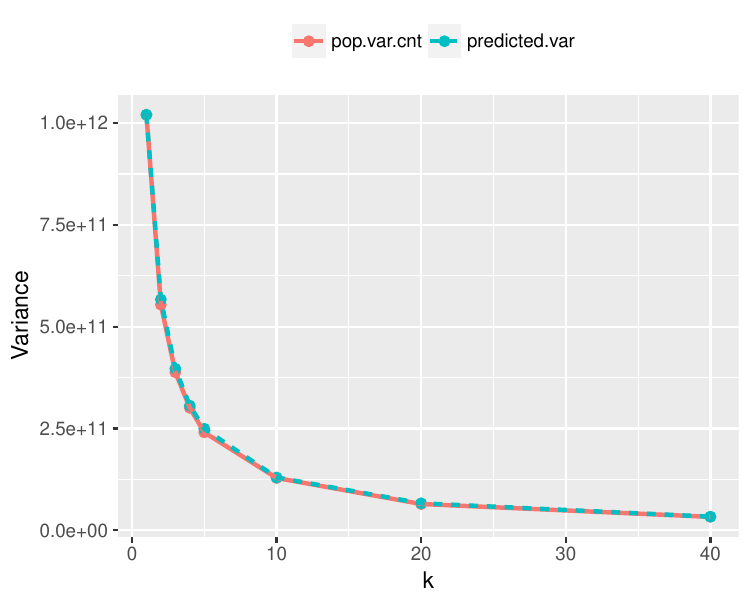}}
   \caption{Eq. \ref{eq:variance-blocks} from Theorem~\ref{thm:variance-blocks} (in blue) versus simulation (in red).}
   \label{fig:var-adj}
\end{figure}

Because $X_i$ shares the same distribution as $V_i$, up to the change of variables $v = 1/r$, the following result follows trivially.

\begin{mythm}
\label{thm:variance_xk}
\begin{equation}
\label{eq:variance}
Var\left[ \frac{1}{k} \sum_{i=1}^k X_i \right] = \frac{(k+1)(2k + 1)}{6k} \left(\frac{1}{r}\right)^2. 
\end{equation}
\end{mythm}

\end{document}